\documentclass[twocolumn,showpacs]{revtex4}
\usepackage{graphicx}

\begin{document}

\title{Behaviour of entanglement and Cooper pairs under relativistic boosts}
\author{Veiko Palge$^{1}$, Vlatko Vedral$^{2,3,4}$, and Jacob A. Dunningham$^{1}$, }
\affiliation{$^{1}$School of Physics and Astronomy, University of Leeds, Leeds LS2 9JT, United Kingdom \\ 
$^{2}$ Clarendon Laboratory, University of Oxford, Parks Road, Oxford OX1 3PU, United Kingdom \\
$^{3}$Centre for Quantum Technologies, National University of Singapore, 3 Science Drive 2, Singapore 117543\\
$^{4}$Department of Physics, National University of Singapore, 2 Science Drive 3, Singapore 117542}

\begin{abstract}
Recent work \cite{Dunningham} has shown how single-particle entangled states are transformed when boosted in relativistic frames for certain restricted geometries. Here we extend that work to consider completely general inertial boosts. We then apply our single particle results to multiparticle entanglements by focussing on Cooper pairs of electrons. We show that a standard Cooper pair state consisting of a spin-singlet acquires spin-triplet components in a relativistically boosted inertial frame, regardless of the geometry. We also show that, if we start with a spin-triplet pair, two out of the three triplet states acquire a singlet component, the size of which depends on the geometry.
This transformation between the different singlet and triplet superconducting pairs may lead to a better understanding of unconventional superconductivity.

\end{abstract}

\maketitle

The theories of quantum mechanics and special relativity form the bedrock of modern physics and are both hugely successful in their own regimes. It is the marriage of these two theories that gives rise to quantum field theory, which is the most accurate description we have of physical reality. Quantum entanglement is often regarded  the fundamental feature that distinguishes quantum and classical physics and so it is important that we are able to describe it using a field theory approach. This requires us to understand how entanglement behaves under relativistic boosts. 

There has already been some work done on how two-particle entanglement changes when viewed from different inertial \cite{Peres2004, Gingrich2002, Alsing2002a,Li2003, Peres2002a} and accelerating \cite{Fuentes2005a} frames. Recent work \cite{Dunningham} has also considered the case of the entanglement between the spin and velocity components of a single particle and shown that single-particle (or mode \cite{Zanardi, Nori2007,Dunningham2}) entanglement depends on the observer but persists right up to the speed of light. However this work only considered the very specific geometry that one of the boosts was orthogonal to both the other boost and to the spin of the particle. In order to draw general conclusions about the nature of this effect, we need to extend the treatment to completely general geometries. An analysis of this is carried out in the first half of the paper.

In the second half of the paper, we show how our single particle results could be used to study the behaviour of multiparticle entangled states under relativistic boosts. For definiteness, we focus on the case of Cooper pairs \cite{BCS1957} but the applicability extends well beyond this. This particular case is interesting because it allows us to understand how superconductivity behaves under boosts when we know that the nature of the Cooper pairs is observer-dependent. In particular, we show that, for general geometries, there is a certain symmetry between spin-singlet and spin-triplet Cooper pairs. Relativistic boosts transform between them and the magnitudes of the different components depend on the geometry of the boosts.

This is interesting because the vast majority of superconductors including high $T_{c}$ cuprates can be explained at the microscopic level by the existence of spin-singlet Cooper pairs. A different class of superconductivity can be attributed to spin-triplet pairs of electrons, but these examples are extremely rare. One such example that has received a lot of  attention since its triplet pairing was experimentally confirmed \cite{Kidwingira2006a} is Sr$_{2}$RuO$_{4}$. This suggests that, among other things, a study of how entanglement behaves under relativistic boosts may give further insight into the nature of exotic superconductors. Although these effects are likely to be very small in most materials where the velocities involved are much smaller than the speed of light, $c$, it is important that they are understood and incorporated into a complete theory. 

Let us start our analysis by considering a massive spin-1/2 particle moving with velocity $v_1$, which is viewed by a relativistic observer traveling at velocity $v_2$. If the two velocities are not collinear, the overall transform is not simply a Lorentz boost, but also involves a rotation. The magnitude of this rotation depends on the values of $v_1$ and $v_2$, and the axis of rotation, $\hat{n} =(n_x,n_y,n_z)$, is given by, $\hat{n} = \hat{v_2}\times\hat{v_1}$.

The unitary matrix representing this rotation was worked out by Wigner in a seminal paper in $1939$ \cite{wigner}. 
This theory is widely known and so we will not repeat it here but just quote the result, however, an excellent treatment can be found in  \cite{Weinberg}. For spin-1/2 particles, the unitary transform that operates on the spin and corresponds to the Wigner rotation is given by,
\begin{equation}
U = (\cos\omega)I  - i\sin\omega\left( n_x \sigma_x  +n_y\sigma_y +n_z\sigma_z\right), \label{rotation}
\end{equation}
where $I$ is the identity operator and $\sigma_x$, $\sigma_y$ and $\sigma_z$ are the Pauli spin operators. The angle of rotation, $\omega$, is given by  \cite{Rhodes}
\begin{eqnarray}
\tan\omega = \frac{\sin\theta}{\cos\theta + D}, \label{tanomega}
\end{eqnarray} 
where $\theta$ is the angle between $v_1$ and $v_2$, 
\begin{eqnarray}
D= \sqrt{\left(\frac{\gamma_1 +1}{\gamma_1 -1}\right)\left(\frac{\gamma_2 +1}{\gamma_2 -1}\right)}
\end{eqnarray} 
and $\gamma_{1,2} = (1-(v_{1,2}/c)^2))^{-1/2}$.

We will now begin by considering how single-particle states are transformed by two boosts and, for definiteness, will consider velocity eigenstates and spin-1/2 particles. For a completely general geometry, the three vectors $v_1$, $v_2$, and spin can each point in any direction. Without loss of generality, we can take the two boosts ($v_1$ and $v_2$) to lie in the $x-y$ plane. Furthermore, we can take $v_1$ to lie along the $y$-axis (see Figure~1). The spin, of course, can point in any direction.  Arranging things in this way, means that the rotation axis points in the $z$-direction and (\ref{rotation}) reduces to the simpler form,
\begin{equation}
U = (\cos\omega)I  - i(\sin\omega) \sigma_z. \label{rotation2}
\end{equation}

\begin{figure}[t] 
\includegraphics[width=7cm]{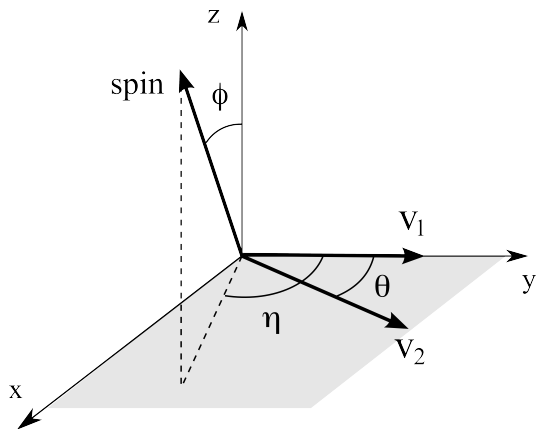}
\caption{Geometry of the boost scheme. A single spin-1/2 particle is given a boost $v_1$ followed by a boost $v_2$. Without loss of generality we take these boosts to define the $x-y$ plane with $v_1$ pointing in the $y$-direction. In general, the particle's spin can point in any direction relative to the boosts. We define the spin direction in terms of the inclination angle $\phi$ and azimuthal angle $\eta$ as shown in the figure.} \label{axesfig}
\end{figure}

The spin vector $|\tilde{\uparrow}\rangle$ and its antiparallel spin $|\tilde{\downarrow}\rangle$ can be written in terms of  up and down spins along the $z$-axis,  $|{\uparrow}\rangle$ and $|{\downarrow}\rangle$ as, 
\begin{eqnarray}
|\tilde{\uparrow}\rangle &=& \cos\frac{\phi}{2}|\uparrow\rangle + ie^{-i\eta}\sin\frac{\phi}{2}|\downarrow\rangle  \label{tildeup}\\
|\tilde{\downarrow}\rangle &=& -\sin\frac{\phi}{2}|\uparrow\rangle + ie^{-i\eta}\cos\frac{\phi}{2}|\downarrow\rangle,\label{tildedown}
\end{eqnarray}
where the angles $\phi$ and $\eta$ are defined in  Fig.~\ref{axesfig}.

By using Eq.~(\ref{tanomega})--(\ref{tildedown}), we can calculate the transform of the initial state
\begin{equation}
(|v_1\rangle + |-v_1\rangle )|\tilde{\uparrow}\rangle 
\end{equation}
due to the boost $v_2$.
Initially there is no entanglement between the velocity and spin parts of the state since they are written as a product. After the boost, the state is
\begin{eqnarray}
\left(\cos\omega_+ -i\sin\omega_+\cos\phi\right)|v_{+}\rangle|\tilde{\uparrow}\rangle &+& i\sin\omega_+\sin\phi |v_{+}\rangle|\tilde{\downarrow}\rangle \nonumber \\
+ \left(\cos\omega_{-} +i\sin\omega_{-}\cos\phi\right)|v_{-}\rangle|\tilde{\uparrow}\rangle &-& i\sin\omega_{-}\sin\phi |v_{-}\rangle|\tilde{\downarrow}\rangle, \nonumber
\end{eqnarray}
where $\omega_{+}$ is given by (\ref{tanomega}), $\omega_{-} = \sin\theta/(-\cos\theta + D)$, and the two velocity eigenstates $v_+$ and $v_{-}$  are the transformations of $v_1$ and $-v_1$ respectively. 

\begin{figure}[b] 
\includegraphics[width=8cm]{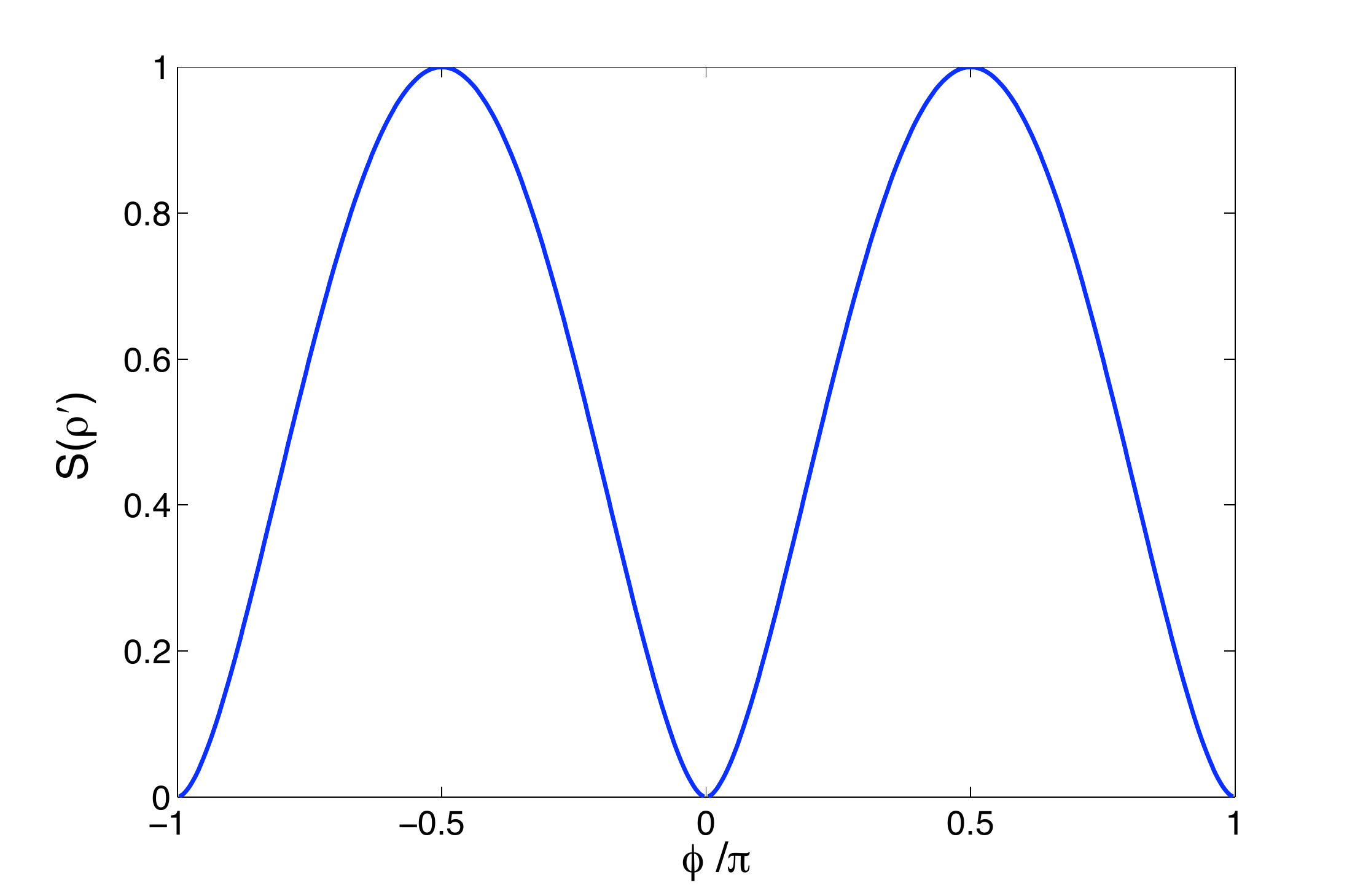}
\caption{Plot of Eq.~(\ref{Sphi}). This shows the entropy of entanglement between the spin and velocity components of a single spin-1/2 particle subjected to two boosts $v_1, v_2 \to c$ as a function of the spin inclination angle $\phi$. } 
\label{entanglement}
\end{figure}

It is clear that the velocity and spin components have now become entangled. We can quantify this by calculating the entropy of entanglement $S(\rho') = -{\rm Tr}\{\rho'\log_2\rho'\}$, where $\rho'$ is the reduced density matrix, found by tracing out the spin component. Taking the limit of $v_1, v_2 \rightarrow c$, we get
\begin{eqnarray}
\rho' &=& \frac{1}{2}\left(|v_+\rangle\langle v_+| + |v_-\rangle\langle v_-|\right)\nonumber \\
&-& \frac{1}{2}\cos\phi\left(|v_+\rangle\langle v_-| - |v_-\rangle\langle v_+|\right),
\end{eqnarray}
which gives the entropy of entanglement as,
\begin{eqnarray}
S(\rho') = 1&-& \frac{1}{2}(1+\cos\phi)\log_2(1+\cos\phi) \nonumber \\
 &-& \frac{1}{2}(1-\cos\phi)\log_2(1-\cos\phi). \label{Sphi}
\end{eqnarray}
This is plotted in Figure~\ref{entanglement}. 
When $\phi =0$, we get $S=0$  (i.e. there is no entanglement between velocity and spin) and when $\phi=\pi/2$, we 
get $S=1$ (i.e. velocity and spin are maximally entangled). In general we see that the entanglement generated by the boost depends on the geometry. Interestingly, in this limit  ($v_1, v_2 \rightarrow c$), the degree of entanglement does not depend on the angle, $\theta$ between the two boosts, but only on the angle, $\phi$.

Now that we have seen how the entanglement of a single spin-1/2 particle is transformed by general boosts, we can use these results to generalize previous work which considered how Cooper pair states are transformed \cite{Dunningham}. The reason for doing this is two-fold. Firstly it allows us to demonstrate by means of a simple example how the transformation of multiparticle states under relativistic boosts can be built up in a straightforward fashion. Secondly, it provides an intriguing insight into superconducting systems.

In standard BCS theory \cite{BCS1957}, superconductivity is explained in terms of the existence of entangled pairs of electrons called Cooper pairs. In conventional superconductors, these electrons form a spin singlet with overall spin zero. However, in more exotic superconductors, they can form spin triplets with overall spin one. One potential issue with superconductors is that they depend on the entangled state but, as we have seen, entanglement depends on the observer. However, there are certain features of superconductivity that we would not expect to be observer-dependent. For example the Meissner effect says that magnet field is expelled from the superconductor  and we would expect this to be true for all observers. Similarly, a magnet is known to levitate over a superconductor due to the induced persistent electric flows on the surface of the superconductor. We would expect all observers to see the magnet levitating. These two observations suggest that a superconductor should appear to be superconducting to observers in all inertial frames. The question is: how do we resolve this with the fact that different observers see different entangled states.

A partial answer was provided in previous work \cite{Dunningham} where it was shown that, for a particular geometry, a singlet Cooper pair acquires triplet components when subjected to relativistic boosts. This suggested that the superconductivity is preserved in different frames but acquires different forms. Here we consider the case of general geometries by direct calculation and will also consider what happens if we boost a spin triplet.
For notational convenience, let us first define the singlet and triplet spin states respectively as,
\begin{eqnarray}
|S\rangle & \equiv&  \frac{1}{\sqrt2} \left( |\tilde{\uparrow},\tilde{\downarrow}\rangle -  |\tilde{\downarrow},\tilde{\uparrow}\rangle\right) \label{singdef} \\
|T_0\rangle & \equiv&  \frac{1}{\sqrt2} \left( |\tilde{\uparrow},\tilde{\downarrow}\rangle +  |\tilde{\downarrow},\tilde{\uparrow}\rangle\right) \label{T0def} \\
|T_\pm\rangle & \equiv& \frac{1}{\sqrt2} \left( |\tilde{\uparrow},\tilde{\uparrow}\rangle \pm  |\tilde{\downarrow},\tilde{\downarrow}\rangle\right). \label{Tpmdef}
\end{eqnarray}
At this point we should issue a warning about our terminology. We will always use the terms singlet and triplet in their nonrelativistic sense as defined in (\ref{singdef})--(\ref{Tpmdef}). This is really an abuse of nomenclature since it is known that the spin operators have a different form in a relativistic frame and therefore so too do the singlet and triplet states. For an excellent discussion of this point see \cite{Caban}. However, it is convenient to retain this terminology here and its meaning should be clear.

Now suppose we start with a singlet (s-wave) Cooper pair,
\begin{equation}
|\Psi\rangle = \frac{1}{\sqrt2}(|v_1, -v_1\rangle + |-v_1, v_1\rangle)|S\rangle,
\end{equation}
and boost in some arbitrary direction $v_2$ as shown in Figure~1. Using the single particle transforms discussed above, the state can be shown to be, 
\begin{eqnarray}
|\Psi\rangle &\rightarrow& \frac{1}{\sqrt{2(1+\Gamma)}} \left\{ (|v_+, v_{-}\rangle + |v_{-}, v_{+}\rangle)|S\rangle \right. \nonumber \\
 && \hspace{-13mm}- \left. i \sqrt{\Gamma} (|v_{+}, v_{-}\rangle - |v_{-}, v_{+}\rangle)  [\sin\phi |T_{-}\rangle 
+ \cos\phi |T_{0}\rangle ] \right\} \label{singlettrans}
\end{eqnarray}
where 
\begin{equation}
\Gamma = \frac{(\gamma_1^2 -1)(\gamma_2^2 -1)}{(\gamma_1 +\gamma_2)^2}\sin\theta.
\end{equation}
The top line of (\ref{singlettrans}) is the just a singlet Cooper pair and the rest of the state is a spin-triplet. So the effect of the boost in any geometry (apart from $\theta=0, \pi$ where the boost are collinear) is to include some triplet components in the state. If we consider the interesting limit $v_1, v_2 \rightarrow c$, we get $\Gamma \to \infty$ and (\ref{singlettrans}) becomes
\begin{eqnarray}
|\Psi\rangle \to \frac{1}{2}(|v_+, v_{-}\rangle - |v_{+}, v_{-}\rangle) [\sin\phi |T_{-}\rangle 
+ \cos\phi |T_{0}\rangle]. \label{singlettrans2}
\end{eqnarray}
In other words the Cooper pair singlet is completely converted into a triplet and this is true independent of the geometry of the boosts (as shown in Figure~1). The relative weighting of the two triplets $|T_{-}\rangle$ and $|T_{0}\rangle$ does, however, depend on the polar angle $\phi$. We should note that, intriguingly, the transformed state when written in the eigenbasis of the Lorentz spin operator defined in \cite{Caban} (as opposed to the nonrelativistic spin operators) remains as a singlet. In other words, such a Lorentz singlet transforms into itself under a Lorentz boost \cite{Caban}.

By our argument that the presence or otherwise of superconductivity should not depend on the observer's frame of reference, it is important that we check how spin-triplet Cooper pairs are transformed when subjected to general boosts. Again using the single particle results discussed above, we can directly calculate the transformations.

Starting with $|T_+\rangle$ we get,
\begin{eqnarray}
&&\frac{1}{\sqrt2}(|v_1, -v_1\rangle - |-v_1, v_1\rangle)|T_+\rangle \longrightarrow \nonumber \\
&&\frac{1}{\sqrt2}(|v_+, v_{-}\rangle - |v_{-}, v_{+}\rangle) [ \cos(\omega_+ - \omega_{-})|T_+\rangle \nonumber \\
&&+i\sin(\omega_{+}-\omega_{-})(\sin\phi|T_0\rangle -\cos\phi |T_{-}\rangle)].
\end{eqnarray}
We see that this state remains as a triplet after the boost. However it acquires different triplet components that depend of the geometry. The other two triplet states transform as follows:
\begin{eqnarray}
&&\hspace*{-7mm}\frac{1}{\sqrt2}(|v_1, -v_1\rangle - |-v_1, v_1\rangle)|T_{-}\rangle \rightarrow \frac{1}{\sqrt2}(|v_+, v_{-}\rangle - |v_{-}, v_{+}\rangle)\nonumber \\
&&\hspace*{-3mm}\times [ \cos(\omega_+)\cos( \omega_{-}) + \sin(\omega_+)\sin( \omega_{-})\cos(2\phi)|T_-\rangle \nonumber \\
&&\hspace*{-3mm} -i\sin(\omega_{+}-\omega_{-})\cos\phi |T_+\rangle - 2\sin\omega_{+}\sin\omega_{-}\sin\phi\cos\phi|T_0\rangle] \nonumber \\
&&\hspace*{-3mm} -\frac{i}{\sqrt2}(|v_+, v_{-}\rangle + |v_{-}, v_{+}\rangle)[\sin(\omega_{+} +\omega_{-})\sin\phi |S\rangle ].
\end{eqnarray}
and finally
\begin{eqnarray}
&&\hspace*{-7mm}\frac{1}{\sqrt2}(|v_1, -v_1\rangle - |-v_1, v_1\rangle)|T_{0}\rangle \rightarrow \frac{1}{\sqrt2}(|v_+, v_{-}\rangle - |v_{-}, v_{+}\rangle)\nonumber \\
&&\hspace*{-3mm}\times [ \cos(\omega_+)\cos( \omega_{-}) - \sin(\omega_+)\sin( \omega_{-})\cos(2\phi)|T_0\rangle \nonumber \\
&&\hspace*{-3mm} +i\sin(\omega_{+}-\omega_{-})\sin\phi |T_+\rangle - 2\sin\omega_{+}\sin\omega_{-}\sin\phi\cos\phi|T_{-}\rangle] \nonumber \\
&&\hspace*{-3mm} - \frac{i}{\sqrt2}(|v_+, v_{-}\rangle + |v_{-}, v_{+}\rangle)[\sin(\omega_{+} +\omega_{-})\cos\phi |S\rangle ].
\end{eqnarray}
Each of these latter two transforms acquires singlet components as well as the other triplet components and the relative size of each of these parts depends on the geometry of the system. This is consistent from what we would expect from the physical argument presented above.

It is also interesting to consider the form of the transforms in the limit $v_1, v_2 \rightarrow c$:
\begin{eqnarray}
&&\hspace*{-5mm}(|v_1, -v_1\rangle - |-v_1, v_1\rangle)|T_{+}\rangle \rightarrow (|v_+, v_{-}\rangle - |v_{-}, v_{+}\rangle)\nonumber \\
&&\hspace*{-3mm} \times [\sin\theta|T_+\rangle +i\cos\theta\cos\phi|T_{-}\rangle - i\cos\theta\sin\phi|T_{0}\rangle],  \label{Tplustrans} 
\end{eqnarray}
\begin{eqnarray}
&&\hspace*{-5mm}(|v_1, -v_1\rangle - |-v_1, v_1\rangle)|T_{-}\rangle \rightarrow (|v_+, v_{-}\rangle - |v_{-}, v_{+}\rangle)\nonumber \\
&&\hspace*{-3mm} \times [\sin\theta\cos^2\phi|T_{-}\rangle +i\cos\theta\cos\phi|T_{+}\rangle +\frac{1}{2}\sin\theta\sin2\phi|T_0\rangle] \nonumber \\
&&\hspace*{-3mm} +(|v_+, v_{-}\rangle + |v_{-}, v_{+}\rangle)[-i\sin\phi |S\rangle], \label{Tminustrans}
\end{eqnarray}
and
\begin{eqnarray}
&&\hspace*{-5mm}(|v_1, -v_1\rangle - |-v_1, v_1\rangle)|T_{0}\rangle \rightarrow (|v_+, v_{-}\rangle - |v_{-}, v_{+}\rangle)\nonumber \\
&&\hspace*{-3mm} \times [\sin\theta\sin^2\phi|T_{0}\rangle -i\cos\theta\sin\phi|T_{+}\rangle -\frac{1}{2}\sin\theta\sin2\phi|T_{-}\rangle] \nonumber \\
&&\hspace*{-3mm} +(|v_+, v_{-}\rangle + |v_{-}, v_{+}\rangle)[-i\cos\phi |S\rangle]. \label{Tzerotrans}
\end{eqnarray}

If we compare these transforms with (\ref{singlettrans2}) for certain cases, there is a pleasing symmetry.
In the case that the spin lies along the $z$-axis, i.e. $\phi = 0$, (\ref{singlettrans2}) gives
\begin{eqnarray}
\frac{1}{\sqrt2}(|v_1, -v_1\rangle + |-v_1, v_1\rangle)|S\rangle \longrightarrow \nonumber \\
\frac{1}{\sqrt2}(|v_+, v_{-}\rangle - |v_{-}, v_{+}\rangle)|T_0\rangle 
\end{eqnarray}
and, ignoring global phases, (\ref{Tzerotrans}) gives
\begin{eqnarray}
(|v_1, -v_1\rangle - |-v_1, v_1\rangle)|T_{0}\rangle \longrightarrow \nonumber \\
\frac{1}{\sqrt2}(|v_{+}, v_{-}\rangle + |v_{-}, v_{+}\rangle)|S\rangle.
\end{eqnarray}
So, starting with the singlet we get the triplet $|T_0\rangle$ and, starting with $|T_0\rangle$, we get the singlet.
Similarly, if we consider the case that the spin lies in the plane defined by the two boosts, i.e. $\phi = \pi/2$, (\ref{singlettrans2}) gives
\begin{eqnarray}
\frac{1}{\sqrt2}(|v_1, -v_1\rangle + |-v_1, v_1\rangle)|S\rangle \longrightarrow \nonumber \\
\frac{1}{\sqrt2}(|v_+, v_{-}\rangle - |v_{-}, v_{+}\rangle)|T_{-}\rangle 
\end{eqnarray}
and, ignoring global phases, (\ref{Tminustrans}) gives
\begin{eqnarray}
\frac{1}{\sqrt2}(|v_1, -v_1\rangle - |-v_1, v_1\rangle)|T_{-}\rangle \longrightarrow \nonumber \\
\frac{1}{\sqrt2}(|v_{+}, v_{-}\rangle + |v_{-}, v_{+}\rangle)|S\rangle.
\end{eqnarray}
So, starting with the singlet we get the triplet $|T_{-}\rangle$ and, starting with $|T_{-}\rangle$, we get the singlet.
As we saw above (\ref{Tplustrans}), the $|T_{+}\rangle$ and singlet states are not coupled by these boosts. We can also see this from (\ref{singlettrans}).

In conclusion, we have studied how the entanglement between the spin and velocity components of a single particle are transformed under the influence of pairs of relativistic boosts in completely general directions. In particular, we have seen that if spin and velocity are initially not entangled, they become entangled and the degree of entanglement depends on the magnitude and relative orientation of the boosts. In the limit that the boosts both approach $c$, the degree of entanglement depends only on the angle that the spin makes with the plane of the two boosts.

We then extended this idea by using the single particle transforms to show how we could calculate the transformation of multiparticle entangled states due to relativistic boosts in general geometries. We focussed in particular on Cooper pair states and found that the singlet state acquires triplet components and that the magnitude of these depend on the geometry of the boosts. In the limit that both boosts approach $c$, the singlet is transformed completely into a triplet regardless of the geometry. Similarly we showed that initial triplet states acquire singlet components under pairs of relativistic boosts. This understanding of how entanglement transforms in relativistic frames may help enhance our understanding of superconductivity as well as providing a valuable tool for developing a general quantum field theory description of entanglement.

{\bf Acknowledgements:}
This work was financially supported by the United Kingdom EPSRC and the European Science Foundation.

\end{document}